\documentclass[sn-mathphys]{sn-jnl}% Math and Physical Sciences Reference Style

\usepackage{amsmath}
\usepackage{amsfonts}
\usepackage{amssymb}
\usepackage{hyperref}
\usepackage{subfigure}
\usepackage{graphicx}% Include figure files
\usepackage{dcolumn}% Align table columns on decimal point
\usepackage{bm}% bold math
\usepackage{lastpage}
\jyear{2021}%

\theoremstyle{thmstyleone}%
\theoremstyle{thmstyletwo}%

\theoremstyle{thmstylethree}%

\begin{document}

\title[Sch.BH perturbed by FFMF]{Schwarzschild black hole perturbed by a force-free magnetic field}

%%=============================================================%%
%% Prefix	-> \pfx{Dr}
%% GivenName	-> \fnm{Joergen W.}
%% Particle	-> \spfx{van der} -> surname prefix
%% FamilyName	-> \sur{Ploeg}
%% Suffix	-> \sfx{IV}
%% NatureName	-> \tanm{Poet Laureate} -> Title after name
%% Degrees	-> \dgr{MSc, PhD}
%% \author*[1,2]{\pfx{Dr} \fnm{Joergen W.} \spfx{van der} \sur{Ploeg} \sfx{IV} \tanm{Poet Laureate}
%%                 \dgr{MSc, PhD}}\email{iauthor@gmail.com}
%%=============================================================%%

%\author[1]{\fnm{Yousef} \sur{Sobouti}}\email{sobouti@iasbs.ac.ir}
%\equalcont{These authors contributed equally to this work.}
\author*[1,2]{\fnm{Haidar} \sur{Sheikhahmadi}}\email{h.sh.ahmadi@gmail.com;h.sheikhahmadi@ipm.ir}

%\author[1,2]{\fnm{Third} \sur{Author}}\email{iiiauthor@gmail.com}
%\equalcont{These authors contributed equally to this work.}
%\affil[1]{ \orgname{Institute for Advance Studies in Basic Sciences (IASBS)}, \orgaddress{\street{Gava Zang}, \city{Zanjan}, \postcode{45137-66731}, \state{Zanjan}, \country{Iran}}}

\affil[1]{\orgdiv{Center for Space Research}, \orgname{North-West University}, \orgaddress{ \city{Mahikeng}, \postcode{2745}, \country{South Africa}}}

\affil[2]{\orgdiv{School of Astronomy}, \orgname{Institute for Research in Fundamental Sciences (IPM)}, \orgaddress{ \city{Tehran}, \postcode{19395-5531}, \country{Iran}}}

%%==================================%%
%% sample for unstructured abstract %%
%%==================================%%

\abstract{
We envisage a  black hole perturbed by a force-free magnetic field (FFMF) outside and attempt to determine its structure. We suppose the metric that describes this black hole is of the static spherical type, that is Schwarzschild, and the energy-momentum tensor emanating from an FFMF source perturbs this background metric, in this regard one can imagine a magnetic accretion disk around the black hole. By solving the equations for such a configuration, we will show that in addition to modifying the diagonal elements of the background metric, we will also see the non-zeroing of the off-diagonal elements of the general metric, one of the immediate consequences of which will be a static to stationary transition.\\
\\
\small{“\emph{Space-time tells matter how to move; matter tells space-time how to curve}”
	\hspace{4cm}\emph{John Archibald Wheeler}} 
}



%%\pacs[JEL Classification]{D8, H51}

%%\pacs[MSC Classification]{35A01, 65L10, 65L12, 65L20, 65L70}

\maketitle

{\section{Introduction}
\label{Introduction}}
Our knowledge of FFMFs  actually originated from the astrophysical phenomenae and their observational data. For example, one can imagine a plasma in which the flow velocity is much less than both the isothermal sound  and Alfv\'{e}n velocities, then the equations governing the evolutions of the plasma  are reduced to magnetohydrodynamics (MHD) \cite{Aschwanden}. In such a configuration, when both gravitational and hydrodynamic forces in comparison to the local magnetic stresses are small, the force governing the MHD evolutions will be of the free-force type \cite{Chandar-Kendlerr,Priest}. Indeed, the philosophy behind this nomenclature  is that the Lorentz force can be ignored, at least up to the first order of the perturbation, because there is no need to confront the electric current force with the other forces in this configuration \cite{Gerald}.
 Therefore, due to the presence of a strong FFMF, comparing to the hydrodynamic energy and thermal motion, and the resulting high energy in the many astronomical phenomenae we know, it is an interesting subject to study more deeply. As a classic and well-known example, for the community, we can take the Sun and its corona \cite{Aschwanden} as an accessible example. The  magnetic field strength on the surface of the sun is clearly not homogeneous and varies from $2000$ to $3000$ $G$, which $G$ stands for Gauss. Of course, for the background of a quiet sun and for its coronal holes, values of $\sim0.1$ to $\sim0.5$ $G$ can be measured by utilizing the spectroscopy of Zeeman splitting lines \cite{Chandrasekhar}.

 One of the cases in which these types of fields are of particular importance is the discussion of accretion disks that surround compact objects such as astrophysical black holes \cite{EHT1}, which could be understood as a consequence of the Blandford-Znajek proposal for assuming that the space around a black hole is filled with plasma \cite{J2}. The physics of this  prototype, the approximate solutions introduced to address satisfactorily the physics behind both the accretion disks and jets, and the phenomenae to which such a configuration applies can be traced in the literature \cite{Ref16J,Ref16Ja,Ref16Jb,Ref16Jc,Ref16Jd,Ref16Je}.
 There are still debates about various solutions, both the exact and the analytical ones, in the presence of a FFMF, and as clearly noted in Jacobson and his collaborators' article \cite{Jac}, except for a few metrics that one can find exact solutions other solutions are all numerical.\\ But here we want to look at the problem in a different way. By finding the values of the magnetic field and consequently the electromagnetic strength tensor  for a Schwarzschild metric,  and considering the emergent energy momentum tensor aiming at perturbing the background metric, one can study the modifications  induced to such a metric. Based on these, our main motivation for conducting this study is to examine how the background metric is modified in the presence of an accretion disk that follows the FFMF equations. Some observations suggest that the black hole in such a configuration should be of the rotating type, and we will investigate whether the disk-induced disturbances are capable of rotating the central body, and if so, what will the off-diagonal coefficients of the modified metric look like?
 \\
 This article is organized as follows:\\
 In the Sec.\ref{Force Free Magnetic Field} we obtain the FFMF equations and electromagnetic strength tensor for a 4-D Cartesian spherical background. In Sec.\ref{Linearized Theory of Gravity} after calculating  the relevant energy-momentum tensor for the  Schwarzschild background metric we  study the perturbed metric with the help of the results of section \ref{Force Free Magnetic Field} and try to obtain the elements that have been arisen through this modification. Then Sec.\ref{Discussion1} ais devoted to discussion and concluding remarks.

{\section{Force-Free  Magnetic Fields, Mathematical Toolkits}
\label{Force Free Magnetic Field}}
In this section we want to work out the answers that can be obtained in the presence of force-free electrodynamics. In doing so,  one first calculates the various components of the FFMF in a spherical Cartesian coordinate, and then he can transform the electromagnetic strength tensors to a Schwarzschild background. There are many examples that can be explored using a free-force electrodynamics, and consequently the Blandford-Znajek mechanism,  including astronomical phenomenae such as disks around massive black holes, e.g. ones at the center of the galaxy M 87, and jets or the gamma ray bursts, GRBs, as well. As already mentioned, although the configuration of the issue is the same, the approach we are considering is somewhat different. Indeed, we are looking at how the presense of the FFMF energy-momentum tensor, which plays the role of a source, changes the  Schwarzschild metric.\\
{\subsection{Spherical Symmetric Solutions to Force-Free Fields }
\label{Force Free Magnetic Field-SubSec-A}}
To begin with, we consider the conditions governing a magnetostatic system. Obviously, neglecting the effects of external forces, the Navier-Stokes equation for a magnetostatic plasma reduces to the following equation \cite{Aschwanden}
\begin{equation}\label{Navier-Stokes}
F=\nabla P = J \times B\,,
\end{equation}
where $F$ and $P$ stand for Lorentz force and thermodynamic pressure respectively, $B$ refers the magnetic field, $J=(\nabla\times B)/\mu$ is the electric current density and $\mu$ the magnetic permeability. For the problem we are considering, usually the plasma pressure  can be ignored, comparing to the magnetic pressure $B^{2}/2\mu$, therefore Eq.\eqref{Navier-Stokes} can be expressed as follows
\begin{equation}\label{J*B}
F=J \times B = 0\,,
\end{equation}
  implies the Lorentz force of the plasma is negligible and the reason behind the force-free naming  is clear here  \cite{Aschwanden}.  In this configuration the electric current density is either zero or parallel to the magnetic field in which the first one is trivial and we consider the latter one indicating that $J$ and $B$ are parallel to each other
  \begin{equation}\label{Parrallel-JtoB}
  \mu J = \alpha B\,,
  \end{equation}
   where $\alpha$ is a scalar function of coordinate. Considering Maxwell's equations
   \begin{equation*}
        \nabla  \times B = \mu J,\,\,\,\,\,\,\,\,\,\,\,\,\,\,\,\nabla  \cdot B = 0\,,
   \end{equation*}
    and after a bit of calculation, we can reach the equation of FFMF as
\begin{equation}\label{2-1}
{\nabla ^2}B + {\alpha ^2}B = 0\,,
\end{equation}
that is Helmholtz equation. Following \cite{Chandar-Kendlerr}, there is a scalar function $\psi$, that satisfies
\begin{equation}\label{3-1}
{\nabla ^2}\psi + {\alpha ^2}\psi = 0\,,
\end{equation}
then one is be able to find three independent solutions namely  solonoidal, toroidal and poloidal respectively as follows
\begin{equation}\label{4-1Fa}
  L = \nabla \psi \,,
\end{equation}
\begin{equation}\label{4-1Fb}
T = \nabla  \times (\hat e\psi )\,,
\end{equation}
and
\begin{equation}\label{5-1F}
S = \frac{1}{\alpha }\nabla  \times T\,,
\end{equation}
where $\hat e$ refers a unit vector in a specific direction  \cite{Aschwanden,Chandar-Kendlerr,Priest}. After some algebra the magnetic field $B$ can be expressed as
\begin{equation}\label{4-1}
B=\frac{1}{\alpha }\nabla  \times \nabla  \times (\hat e\psi ) + \nabla  \times (\hat e\psi )\,,
\end{equation}
 where obviously satisfies Eq.\eqref{3-1}.  Now by considering a Cartesian  spherical metric
 \begin{equation}\label{5-1}
d{S^2} = d{t^2} - d{r^2} - {r^2}d{\theta ^2} -{r^2}{\sin ^2}\theta d{\varphi ^2}\,,
\end{equation}\label{5-1Fa}
  by separating the variables for $\psi(r, \theta)=R(r)\Theta(\theta)$ in Eq.(\ref{3-1}) and by virtue of Eq.(\ref{5-1}), one can obtain a solution for $\psi$ as
  \begin{equation}\label{5-1Fb}
    \psi=(\kappa r) j_{n}(\kappa r)(1-\cos \theta) P_{n}^{(1,-1)}(\cos \theta)\,,
  \end{equation}
  where $ j_{n}$ refers spherical Bessel function and $P_{n}^{(1,-1)}$ is the Jacobi polynomials and $\kappa=\alpha\left(C^{2}+\psi^{2}\right)^{1 / 2}/\psi $ in which $C$ is a constant, assuming  $C=0$ then $\alpha$  is reduced to $\kappa$,  \cite{Gerald}.
% \begin{eqnarray}\label{7-1}
%\frac{1}{{{r^2}}}\frac{d}{{dr}}\big({{{r^2}}}\frac{{dR(r)}}{{dr}}\big) + {\alpha ^2}R(r) - \frac{{n(n + 1)}}{{{r^2}}}R(r)&=0 ,\cr
%\frac{1}{{\sin \theta }}\frac{d}{{d\theta }}\big(\sin \theta \frac{{d\Theta (\theta )}}{{d\theta }}\big) + n(n + 1)\Theta (\theta ) &=0 .
%\end{eqnarray}
Considering Eq.\eqref{5-1Fb} for $n=1$ results
 \begin{equation}\label{8-1}
\psi(r, \theta)=\Big( \frac{{\sin (\kappa r)}}{{{{\kappa r}}}} - {{\cos (\kappa r)}}\Big) \sin \theta\,.
\end{equation}
Now one can obtain different components of the axial symmetric magnetic field, independent of $\phi$, as follows
\begin{equation}\label{MagneticBr}
  {B_r} = \frac{1}{{{r^2}\sin \theta }}{\partial _\theta }\psi\,,
\end{equation}
\begin{equation}\label{MagneticBtheta}
{B_\theta } =  - \frac{1}{{r\sin \theta }}{\partial _r}\psi\,,
\end{equation}
and
\begin{equation}\label{MagneticPhi}
  {B_\phi } = \frac{1}{{r\sin \theta }}f(\psi )\,,
\end{equation}
where $f(\psi)$ is defined as $\kappa \psi$.
Considering these results and by considering Eq.\eqref{5-1} the covariant electromagnetic strength tensor for FFMF can be expressed as,
\begin{equation}\label{Falhpabeta}
  {F_{\delta \rho }} = \left( {\begin{array}{*{20}{c}}
0&{{E_r}}&{r{E_\theta }}&{r\sin \theta {E_\phi }}\\
{ - {E_r}}&0&{ - r{B_\phi }}&{r\sin \theta {B_\theta }}\\
{ - r{E_\theta }}&{r{B_\phi }}&0&{ - {r^2}\sin \theta {B_r}}\\
{ - r\sin \theta {E_\phi }}&{ - r\sin \theta {B_\theta }}&{{r^2}\sin \theta {B_r}}&0
\end{array}} \right)\,,
\end{equation}
and immediately the contravariant tensor can be obtained as
\begin{equation}\label{FCont-alhpabeta}
  {F^{\delta \rho }} =\left( {\begin{array}{*{20}{c}}
0&{ - {E_r}}&{ - \frac{{{E_\theta }}}{r}}&{ - \frac{{{E_\phi }}}{{r\sin \theta }}}\\
{{E_r}}&0&{ - \frac{{{B_\phi }}}{r}}&{\frac{{{B_\theta }}}{{r\sin \theta }}}\\
{\frac{{{E_\theta }}}{r}}&{\frac{{{B_\phi }}}{r}}&0&{ - \frac{{{B_r}}}{{{r^2}\sin \theta }}}\\
{\frac{{{E_\phi }}}{{r\sin \theta }}}&{ - \frac{{{B_\theta }}}{{r\sin \theta }}}&{\frac{{{B_r}}}{{{r^2}\sin \theta }}}&0
\end{array}} \right)\,,
\end{equation}
where ${E_r} = {E_\theta } = {E_\phi } = 0$, \cite{Jac,Blinder:2004ik}.\\
But that is not the whole point. With the help of these results, we want to study the problem of perturbation induced by a FFMF to a Schwarzschild metric; in doing so we will go through the problem  in the next section.\\

%%%%%%%%%%%%%%%%%%%%%%%%%%%%%%%%%%%%%%%%%%%%%%%%%%%%%%%%%%%%%%%%%%%%%%%%%%%%%%%%%%%%%%%%%%
%%%%%%%%%%
%%%%%%%%%%%%%%%%%%%%%%%%
%%%%%%%%%%%%%%%%%%%%%%%%%%%%%%%%%%%%%%%%%%%%%%%%%%%%%%%%%%%%%%%%%%%%%%%%%%%%%%%%%%%%%%%%%%
%%%%%%%%%%
%%%%%%%%%%%%%%%%%%%%%%
%%%%%%%%%%%%%%%%%%%%%%%%%%%%%%%%%%%%%%%%%%%%%%%%%%%%%%%%%%%%%%%%%%%%%%%%%
%%%%%%%%%%%%%%%%%%%%%%%%%%%%%%%%%%%%%%%%%%%%%%%%%%%%%%%%%%%%%%%%%%%%%%%%%%
%%%%%%%%%%%%%%%%%%%%%%%%%%%%%%%%%%

%%%%%%%%%%%%%%%%%%%%%%%%%%%%%%%%%%%%%%%%%%%%%%%%%%%%%%%%%%%%%%%%%%%%%%%%%
%%%%%%%%%%%%%%%%%%%%%%%%%%%%%%%%%%%%%%%%%%%%%%%%%%%%%%%%%%%%%%%%%%%%%%%%%%

%%%%%%%%%%%%%%%%%%%%%%%%%%%%%%%%%%%%%%%%%%%%%%%%%%%%%%%%%%%%%%%%%%%%%%%%%%%%%%%%%%%%%%%%%%%%%%
%%%%%%%%%\section{Linearized Theory of Gravity}%%%%%%%%%%%%%%%%%%%
%%%%%%%%%%%%%%%%%%%%%%%%%%%%%%%%%%%%%%%%%%%%%%%%%%%%%%%%%%%%%%%%%%%%%%%%%%%%%%%%%%%%%%%%%%%%%%

{\section{Linearized Theory of Gravity}
\label{Linearized Theory of Gravity}}
In this stage, we want to consider the  linearized theory of gravity \cite{Weinberg,MTW,Padabnamaha,Ryder}. In such a theory one can  write down the general metric as
\begin{equation}\label{1-0}
g_{\mu\nu}=g_{\mu\nu}^{(0)}+h_{\mu\nu},
\end{equation}
where
\begin{equation}\label{Schwarzschil01}
g_{\mu\nu}^{(0)} = \left( (1 - \frac{{2 G M}}{r}), -{(1 - \frac{{2 G M}}{r})^{-1}},- {r^2}, -{r^2}{\sin ^2}\theta\right)\,,
\end{equation}
 stands for Schwarzschild metric, that is background metric, and $h_{\mu\nu}$ refers the perturbations subjected to the presence of a FFMF source.
 We want to examine the equations that govern evolutions of the system in the aforementioned configuration. That is, we analyze how a system with a Schwarzschild metric in the background due to the presence of force-free electrodynamics  is disturbed.
 In such analysis, it is better to introduce two parameters related to the scales that perturbations vary $\lambda$, as well as the scales at which the background metric varies, $L$. Such an assumption will dramatically help  in simplifying the calculations and will have interesting results.
 For the metric introduced in the Eq.\eqref{1-0}  the first-order Christoffel symbols of the perturbation can be obtained in the following form
 \begin{equation}\label{Christoffel 01}
\Gamma_{\mu\nu}^{(1)\eta}=\frac{1}{2}\left(\nabla_{\nu} h_{\mu}^{\eta}+\nabla_{\mu} h_{\nu}^{\eta}-\nabla^{\eta} h_{\nu \mu}\right)\,.
\end{equation}
 Then with  the help of this definition, we arrive at the following relation for Riemann tensor
% \begin{widetext}
 \begin{equation}\label{Riemann01}
\begin{gathered}
R_{\mu\nu\epsilon}^{(1)\eta}=\frac{1}{2}\left(\nabla_{\nu} \nabla_{\epsilon} h_{\mu}^{\eta}+\nabla_{\nu} \nabla_{\mu} h_{\epsilon}^{\eta}-\nabla_{\nu} \nabla^{\eta} h_{\mu \epsilon}-\nabla_{\epsilon} \nabla_{\nu} h_{\mu}^{\eta}\right. \\
\left.-\nabla_{\epsilon} \nabla_{\mu} h_{\nu}^{\eta}+\nabla_{\epsilon} \nabla^{\eta} h_{\mu \nu}\right)\,,
\end{gathered}
\end{equation}
%\end{widetext}
 and according to the normal process in which one can contract the Riemann tensor, he can reach the Ricci tensor in the following relation
%  \begin{widetext}
 \begin{equation}\label{Ricci01}
R_{\mu\nu}^{(1)}=R_{\mu \eta \nu}^{(1) \eta}=\frac{1}{2}\left(\nabla_{\eta} \nabla_{\nu} h_{\mu}^{\eta}+\nabla_{\eta} \nabla_{\mu} h_{\nu}^{\eta}-\nabla_{\eta} \nabla^{\eta} h_{\mu k}-\nabla_{\nu} \nabla_{\mu} h\right)\,,
\end{equation}
%\end{widetext}
 where $h$ is the trace of $h_{\mu\nu}$. The Ricci scalar can also be expressed, with the help of the relations obtained to determine the components of the mixed Ricci tensor, as
 \begin{equation}\label{RicciScalar01}
R_{\nu}^{(1)\mu}=g^{\mu \eta(0)} R_{\nu \eta}^{(1)}-h^{\mu \eta} R_{\nu \eta}^{(0)}\,.
\end{equation}
 At this stage, for the sake of convenience it is better to use the trace reversed metric, i.e. $\bar{h}_{\mu\nu}$, which  can be expressed in the following form
 \begin{equation}\label{hbar01}
\bar{h}_{\mu \nu} \equiv h_{\mu \nu}-(\frac{1}{2}) g_{\mu \nu}^{(0)} h\,.
\end{equation}
 Now let's make a very small coordinate transformation, i.e. $x^\mu \longrightarrow x^\mu+\xi^\mu$ and from the Lorentz gauge condition, that ends in the following equation
 \begin{equation}\nonumber
\square \xi^{\mu}=-\nabla_{\nu} \bar{h}_{\mu^{\nu}}^{\nu}\,,
\end{equation}
 we can get a very important equation for the propagation of an electromagnetic wave on a curved background as
 \begin{equation}\label{wavepropagation01}
\square \bar{h}_{\mu \nu}+2 R_{\eta \mu \epsilon \nu}^{(0)} \bar{h}^{\eta \epsilon}=4G \mathcal{T}_{\mu\nu}\,,
\end{equation}
 \cite{Padabnamaha}, where
 \begin{equation}\label{E-M01}
   {\mathcal{T}_{\mu \nu }} = {F_{\mu \gamma }}F_\nu ^\gamma  - \frac{1}{4}g_{\mu\nu}^{(0)}({F_{\sigma \eta }}{F^{\sigma \eta }})\,,
 \end{equation}
is the energy-momentum tensor   \cite{Padabnamaha,Jackson}. In our study, the physical conditions prevail in such a way that we can use up to the order of $\mathcal{O}(\lambda^2/L^2)$ that causes the omission of the second term in the equation \eqref{wavepropagation01} and finally achieve a simpler form below
  \begin{equation}\label{wavepropagation02}
\square \bar{h}_{\mu \nu}=4G \mathcal{T}_{\mu\nu}\,.
\end{equation}
 This equation allows us to use the radiative wave form and define the following relation
 \begin{equation}\label{Waveform02}
{{\bar h}_{\mu \nu {\kern 1pt} }}(r) = 4 G \int {\frac{{{\mathcal{T}_{\mu \nu }}(r)}}{{ \mid\mathbf{r}-\mathbf{R}\mid}}} {d^3}R.
\end{equation}
 where, due to magnetostatic circumstances, one can define, see  \cite{Padabnamaha,Ryder}
 \begin{equation}\label{Inverse r-R}
\begin{aligned}
\mid\mathbf{r}-\mathbf{R}\mid &=\left(r^{2}-2 \mathbf{r} \cdot \mathbf{R}+R^{2}\right)^{1 / 2} \approx r\left(1-\frac{\mathbf{r} \cdot \mathbf{R}}{r^{2}}\right)+\cdots\,, \\
\mid\mathbf{r}-\mathbf{R}\mid^{-1} &=\left(r^{2}-2 \mathbf{r} \cdot \mathbf{R}+R^{2}\right)^{-1 / 2} \approx \frac{1}{r}\left(1-\frac{\mathbf{r} \cdot \mathbf{R}}{r^{2}}\right)+\cdots\,.
\end{aligned}
\end{equation}
 To determine the  energy-momentum tensor and consequently the various components of the perturbations, it is necessary to obtain the covariant and contravariant electromagnetic strength tensors for the Schwarzschild background. By utilizing the tensor transformation rules and using Eqs. \eqref{MagneticBr} to \eqref{FCont-alhpabeta} and \eqref{Schwarzschil01} they read respectively

  \begin{equation}\label{StrenghSch01}
 {F_{\mu \nu }} = \left( {\begin{array}{*{20}{c}}
0&0&0&0\\
0&0&{ - \frac{{{B_\phi }}}{{r - 2GM}}}&{\frac{{\sin \theta {B_\theta }}}{{r - 2GM}}}\\
0&{\frac{{{B_\phi }}}{{r - 2GM}}}&0&{ - {r^2}\sin \theta {B_r}}\\
0&{ - \frac{{\sin \theta {B_\theta }}}{{r - 2GM}}}&{{r^2}\sin \theta {B_r}}&0
\end{array}} \right)\,,
\end{equation}

 and
  \begin{equation}\label{StrenghSch02}
 {F^{\mu \nu }} = \left( {\begin{array}{*{20}{c}}
0&0&0&0\\
0&0&{ - (r - 2GM){B_\phi }}&{\frac{{(r - 2GM){B_\theta }}}{{\sin \theta }}}\\
0&{(r - 2GM){B_\phi }}&0&{ - \frac{{{B_r}}}{{{r^2}\sin \theta }}}\\
0&{\frac{{(r - 2GM){B_\theta }}}{{\sin \theta }}}&{\frac{{{B_r}}}{{{r^2}\sin \theta }}}&0
\end{array}} \right)\,.
\end{equation}
Now we are in a situation where we can calculate the various components of the energy-momentum tensor, \eqref{E-M01},  with the help of equations \eqref{StrenghSch01} and \eqref{StrenghSch02} as follows, the diagonal componets,
\begin{equation}\label{T00}
  {\mathcal{T}_{00}} = \frac{{ - (r - 2GM)}}{{2r}}(B_r^2 + B_\theta ^2 + B_\phi ^2)\,,
\end{equation}
\begin{equation}\label{T11}
  {\mathcal{T}_{11}} = \frac{r}{{2(r - 2GM)}}(B_\theta ^2 + B_\phi ^2-B_r^2)\,,
\end{equation}
\begin{equation}\label{T22}
  {\mathcal{T}_{22}} = \frac{{{r^2}}}{2}(B_r^2 + B_\phi ^2-B_\theta ^2)\,,
\end{equation}

\begin{equation}\label{T33}
  {\mathcal{T}_{33}} = \frac{{{r^2}{{{\mathop{\rm Sin}\nolimits} }^2}\theta }}{2}(B_r^2 + B_\theta ^2-B_\phi ^2)\,,
\end{equation}
and the off-diagonal elements can be expressed as
\begin{equation}\label{T12}
  {\mathcal{T}_{12}} = - {r^2}(B_\phi ^2)\,,
\end{equation}
\begin{equation}\label{T13}
  {\mathcal{T}_{13}} =  - {r^2}{{\mathop{\rm Sin}\nolimits} ^2}\theta (B_\theta ^2)\,,
\end{equation}
\begin{equation}\label{T21}
  {\mathcal{T}_{21}} = - \frac{r}{{(r - 2GM)}}(B_\phi ^2)\,,
\end{equation}
\begin{equation}\label{T31}
  {\mathcal{T}_{31}} = - \frac{r}{{(r - 2GM)}}(B_\theta ^2)\,,
\end{equation}
\begin{equation}\label{T23}
  {\mathcal{T}_{23}} =  - {r^2}{{\mathop{\rm Sin}\nolimits} ^2}\theta (B_r^2)\,,
\end{equation}
\begin{equation}\label{T32}
  {\mathcal{T}_{32}} =   - {r^2}(B_r^2)\,,
\end{equation}
where
\begin{eqnarray}\label{Brthetaphi}\nonumber
B_r^2& =& \frac{{4{{{\mathop{\rm Cos}\nolimits} }^2}(\theta ){{{\mathop{\rm Cos}\nolimits} }^2}(\kappa r)}}{{{r^4}}};\,\,B_\theta ^2 = \frac{{{\kappa ^2}{{{\mathop{\rm Sin}\nolimits} }^2}(\theta ){{{\mathop{\rm Sin}\nolimits} }^2}(\kappa r)}}{{{r^2}}};\\
\,B_\phi ^2 &=& \frac{{{\kappa ^2}{{{\mathop{\rm Sin}\nolimits} }^2}(\theta ){{{\mathop{\rm Cos}\nolimits} }^2}(\kappa r)}}{{{r^2}}}\,,
\end{eqnarray}
and as mentioned above aiming at calculating these equations  one can assume
\[\psi  = \left( {\frac{{\sin \kappa r}}{{\kappa r}} - \cos \kappa r} \right){\sin ^2}\theta \,\,(r \gg 1)\,\Rightarrow\psi  =  - \cos (\kappa r){\sin ^2}\theta\,. \]
Therefore introducing Eqs. \eqref{T00}-\eqref{T32} into Eq. \eqref{Waveform02} one gets
\begin{eqnarray}\label{hbar00}\nonumber
  \bar{h}_{00}&=&\\\nonumber
  &-&\frac{8 \pi  G }{3 k r}\Bigg[-4 G k M \text{Ci}(2 k r)\\\nonumber
  &+&2 \left(k^3+k\right) \big(r-2 G M \log (r)\big)+\sin (2 k r)\Bigg]\,,
\end{eqnarray}
\begin{eqnarray}\label{hbar11}\nonumber
\bar{h}_{11}=-\frac{16 \pi  G  }{3 r^3}\Bigg[2 k r \big\{G M \big(k r (2 \text{Ci}(2 k r)+\log (r))-\sin (2 k r)\big)\\
+r \text{Si}(2 k r)\big\}+(G M+r) \cos (2 k r)+G M+k^2 r^3+r\Bigg]\,,
\end{eqnarray}
\begin{equation}\label{hbar22}
  \bar{h}_{22}=-\frac{4 \pi  G \Big[\left(2 k^2 r^2+1\right) \sin (2 k r)+2 k r (\cos (2 k r)+2)\Big]}{3 k r}\,,
\end{equation}

\begin{eqnarray}\label{hbar33}\nonumber
  \bar{h}_{33}&=&\frac{\pi  G }{720 k r}\Bigg[4 k r \Big((45 \pi -128) k^2 r^2-384\Big)\\\nonumber
&+&3 \Big(2 (128+45 \pi ) k^2 r^2-45 \pi -384\Big) \sin (2 k r)\\
&+&6 (128+45 \pi ) k r \cos (2 k r)\Bigg]\,,
\end{eqnarray}
where $\text{Ci}(2 k r)$ and $\text{Si}(2 k r)$ are Cosine and Sine integrals respectively and both of them vanish at $r\gg 1$. Then by virtue of Eq.\eqref{hbar01} it is easy to obtain
 \begin{equation}\label{hbamunu}
{h}_{\mu \nu} = \bar{h}_{\mu \nu}-(\frac{1}{2}) g_{\mu \nu}^{(0)} \bar{h}\,,
\end{equation}
and considering this fact that the off-diagonal components of the background metric are equal to zero then one has

\begin{equation}\label{hbar12}
  \bar{h}_{12}=h_{12}=\frac{32 \pi  G k^2 \Bigg[\frac{\left(2 k^2 r^2-1\right) \sin (2 k r)}{8 k^3}+\frac{r \cos (2 k r)}{4 k^2}+\frac{r^3}{6}\Bigg]}{3 r}\,,
\end{equation}

 \begin{equation}\label{hbar13}
  \bar{h}_{13}=h_{13}=\frac{128 \pi  G k^2 \Bigg[-\frac{\left(2 k^2 r^2-1\right) \sin (2 k r)}{8 k^3}-\frac{r \cos (2 k r)}{4 k^2}+\frac{r^3}{6}\Bigg]}{15 r}\,,
\end{equation}

 \begin{eqnarray}\label{hbar21}\nonumber
  \bar{h}_{21}&=&h_{21}=\frac{32 \pi  G}{3 r} \Bigg[G k^2 M \text{Ci}(2 k r)+G k^2 M \log (r)\\
 & +&\frac{k^2 r}{2}+\frac{1}{4} k \sin (2 k r)\Bigg]\,,
\end{eqnarray}

\begin{eqnarray}\label{hbar31}\nonumber
  \bar{h}_{31}&=&h_{31}=\frac{32 \pi  G }{3 r}\Bigg[-G k^2 M \text{Ci}(2 k r)+G k^2 M \log (r)\\
  &+&\frac{k^2 r}{2}-\frac{1}{4} k \sin (2 k r)\Bigg]\,,
\end{eqnarray}

\begin{equation}\label{hbar23}
  \bar{h}_{23}=h_{23}=\frac{128 \pi  G \left(\frac{\sin (2 k r)}{4 k}+\frac{r}{2}\right)}{15 r}\,,
\end{equation}

\begin{equation}\label{hbar32}
  \bar{h}_{32}=h_{32}=\frac{64 \pi  G \left(\frac{\sin (2 k r)}{4 k}+\frac{r}{2}\right)}{3 r}\,,
\end{equation}

Here by considering the definition, $\bar{h}=g^{\mu\nu (0)}\bar{h}_{\mu\nu}$  one can obtain trace as
\begin{eqnarray}\label{tracehbar01}\nonumber
%\begin{aligned}
\bar{h}&=&-h=\\\nonumber
&-&\frac{16 \pi  G\left(1-\frac{2 G M}{r}\right) }{3 r^3}\Bigg[2 k r \Big(G M (k r (2 \text{Ci}(2 k r)+\log (r))\\\nonumber
 &-&\sin (2 k r))+r \text{Si}(2 k r)\Big)+(G M+r) \cos (2 k r)+G M\\\nonumber
 &+&k^2 r^3+r\Bigg]+\frac{\pi  G }{720  k r^3 \text{Sin}^2\theta }\Bigg\{4 k r \Big((45 \pi -128) k^2 r^2-384\Big)\\\nonumber
&+&3 \Big(2 (128+45 \pi ) k^2 r^2-45 \pi -384\Big) \sin (2 k r)\\\nonumber
&+&6 (128+45 \pi ) k r \cos (2 k r)\Bigg\}-\frac{4 \pi  G }{3 k r^3}\Bigg[\big(2 k^2 r^2+1\big) \sin (2 k r)\\\nonumber
&+&2 k r \big(\cos (2 k r)+2\big)\Bigg]-\frac{8 \pi  G \left(\frac{2 G M}{r}+1\right) }{3 k r}\Bigg\{+\sin (2 k r)\\\nonumber
&-&4 G k M \text{Ci}(2 k r)+2 \big(k^3+k\big) \big(r-2 G M \log (r)\big)\Bigg\}\,.
%\end{aligned}
\end{eqnarray}

{\section{Discussion and Concluding Remarks}\label{Discussion1}}
Considering the configurations that led to the Reissner-Nordstr\"{o}m \cite{Reissner(1916)} or the Kerr-Newman metric \cite{Newman:1965my}, we tried to test such a possibility for a system in which an electromagnetic plasma plays the role of a source of perturbations.
The interaction  between geometry and matter plays an important role in determining the type of space-time geometric structure. This concept led us to study the effects of the presence of a FFMF source on how the background geometry develops.
 In this paper, it is shown that although the background metric is a diagonal static metric, the developed metric,  in addition to modifying the diagonal elements, will also have non-zero off-diagonal expressions, and can be considered as a static to a stationary transition.\\
In the other words, assuming an accretion disk around a Schwarzschild black hole that obeys FFMF laws, we investigated the effects of such a configuration on changes in background geometry. Actually,  instead of assuming a definite metric and solving force-field equations for it, our goal was to use the technique of the first-order perturbations in a curved background to look at how geometry changes from this perspective. For this purpose, we first obtained the equations of electromagnetic field and its strength tensor for the background metric. Then, with the help of radiation solution and its connection with the  energy-momentum tensor, we were able to extract the modified perturbation components. Interestingly, due to the presence of such sources, the new black hole is no longer static, but has become a stationary black hole due to the non-zeroing of off-diagonal components.\\
But this is not the end of the story, and to examine this new metric in more detail, we need to test them with observational data. The effects of such a source on changes in other metrics can also be examined.
All of these topics could be the subject of future research.\\
\\

{\section*{Acknowledgments}}

H.S. is grateful to Y. Sobouti who basically taught him the physics of the FFMFs.
\\
\\

\end{document}